\documentstyle[12pt,a4,psfig,amssymb]{article}
\newcommand{\smdag}{\mbox {\tiny \dag}}

\newcommand{\vI}{V_{\mbox {\scriptsize I}}}

\begin{document}

\begin{center}
{\Large {\bf
Electromagnetic form factors of the nucleon in a  
relativistic  quark pair creation model}}
\end{center}

\vspace{1cm}

\begin{center}
{\large F. Cano$^{1,2}$, B. Desplanques$^{3}$, P. Gonz\'alez$^{4,5}$, 
S. Noguera$^{4}$}
\end{center}

\begin{center}
$^{1}$Universit\`a degli Studi di Trento, \\
I-38050 Povo (Trento), Italy  \\
\vspace{0.25cm}
$^{2}$CEA-Saclay, DAPNIA/SPhN \\
F-91191 Gif-Sur-Yvette Cedex, France \\
\vspace{0.25cm}
$^{3}$ Institut des Scieces Nucl\'eaires \\
F-38026 Grenoble Cedex, France \\
\vspace{0.25cm}
$^{4}$Departamento de F\'{\i}sica Te\'orica, Universidad de Valencia\\
46100 Burjassot (Valencia), Spain \\
\vspace{0.25cm}
$^{5}$IFIC, Centro Mixto Universidad de Valencia-CSIC, Valencia, Spain

\end{center}

\vspace{1cm}

\begin{abstract}
{\small We study the effects of the $| qqq \; q\bar{q}\rangle$
component of the hadronic wave function on the description of the
electromagnetic structure of the nucleon. Starting with a $3q$
baryonic wave function which describes the baryonic and mesonic low
energy spectrum, the extra $q\bar{q}$ pair is generated through a
relativistic version of the $^3P_0$ model. It is shown that this
model leads to a renormalization of the quark mass that allows one to
construct a conserved electromagnetic current. We conclude that these
dynamical relativistic corrections play an important role 
in reproducing the $Q^2$ dependence of the electromagnetic 
form factors at low $Q^2$.}
\end{abstract}

\vfill
\vspace{1cm}
\noindent 
{\bf PACS}: 13.40.Gp, 13.40Em, 12.39.Ki \\
{\bf Keywords:} Non-relativistic quark models, electromagnetic form
factors, meson cloud.

\vspace{1cm}
\noindent 
fcano@cea.fr \\
desplanq@isn.in2p3.fr \\
Pedro.Gonzalez@uv.es \\
Santiago.Noguera@uv.es

\vfill
 
\newpage

\section{Introduction}

        Electromagnetic processes constitute a basic tool to
investigate the baryon structure since the photon couples to the spin
and flavor of the constituent quarks, revealing their spin-flavor
correlations inside the baryons. This explains  the current
experimental effort along this line (MAMI, ELSA, GRAAL) with specific
experimental programs in TJNAF \cite{BURKERT96}.

        From a theoretical point of view, most analyses rely on the
use of the non-relativistic quark model
\cite{KONIUK80} 
in spite of the fact that for the low-lying
non-strange resonances the velocity of the quarks inside the baryons
may be close to $c$.  Incorporation of two--body exchange currents does
not mean much improvement on the results 
\cite{ROBSON93}.  On the other hand,
attempts to use relativized quark models combined with consistent
transition operators have been carried out
\cite{WARNS90,DONG99} and  
Light-Front and Point--Form 
studies have also been done \cite{KONEN90,WAGENBRUNN00} 
partially solving some of the failures of the non-relativistic approach.
However, a complete understanding of the relevant ingredients in the
description of electromagnetic processes has not been reached yet.
 
        Our aim in this article is to investigate the role played by
some relativistic corrections to the electromagnetic transition
operators, specifically those ones related to the coupling of the
photon to $q\bar{q}$ components of the baryon (mesonic cloud), also
underlying the well known Vector--Meson Dominance phenomenology. The
need for the {\it explicit} contribution of the cloud to describe
electromagnetic interactions of baryons was also concluded in
\cite{CANO00}, where it was shown on very general grounds that meson
exchange in the $qq$ potential can not play the role of the $|qqq \;
q\bar{q} \rangle$ configurations in the baryon. The importance of the
explicit consideration of the meson cloud for electromagnetic
processes has been recently studied in
Ref. \cite{DONG99}. 

        A main motivation for this study comes from the analysis of
strong pionic decay processes where the implementation of the coupling
of the pion to $q\bar{q}$ baryon components through a $^3P_0$ operator
allows a reasonable description of the decay widths \cite{CANO96}. In
comparison to the elementary emission model, the improvement is
especially spectacular for the Roper resonance since the  decay
width for the $N \pi$ channel has changed from a few MeV to few
hundreds MeV.
However, as
shown in \cite{THEUSSL00}, the precise value is open
to discussion.

        By proceeding in the same way for the electromagnetic
transition operator a first simplified model for the  photo and
electroproduction amplitudes of $N(1440)$ was presented in 
\cite{CANO98}. These results suggest that the explicit contribution of
the baryon mesonic cloud (taken implicitly into account in the baryon
spectrum through the effective parameters and/or interactions of the
potential) is an essential ingredient for the description of
transition processes from a non-relativistic quark model scheme.

        Here we apply the same ideas to construct a more complete and
consistent model to deal with electromagnetic processes that we shall
test by evaluating the nucleon form factors. 
We put the
emphasis in the construction of an effective transition operator to be
sandwiched between effective quark--core wave functions as the ones
provided by spectroscopic models. 
In particular we shall center in a model  previously
used to fit the baryon and meson spectrum \cite{SILVESTRE85} and to predict strong pionic decay
widths \cite{CANO96}, though our treatment can be applied to {\it any
other} quark--core model of the baryon structure.

        The paper is organized as follows. In section 2 we discuss 
the direct quark--photon coupling through the elementary emission model
that we shall apply to the calculation of nucleon form
factors. In section 3 we study dynamical relativistic corrections
induced by $q\bar{q}$ pairs in the baryonic medium. A $^3P_0$
model will be used in order to implement the relevant $qqq \;
q\bar{q}$ baryon components. From the consideration of resonant and 
non-resonant diagrams we are driven in section 4 to develop a gauge
invariant model. Results are presented and discussed in section 5.

\section {The elementary emission model (EEM)}

        In the EEM the baryon transition process $B \rightarrow
B'\gamma$ is described by assuming that the photon is emitted by a
constituent quark of the baryon (Fig. 1a). 
The relevant matrix element between quark states is written as:

\begin{equation} \langle q(\vec{p}\:') \gamma(\vec{q},\lambda)|
H_{qq\gamma} | q(\vec{p}\,) \rangle = \frac{e_q}{(2 \pi)^{3/2}}
\frac{1}{(2 \omega_\gamma)^{1/2}} \delta^{(3)}(\vec{p} - \vec{p}\;'
- \vec{q} \, ) O_{qq\gamma}^{\mbox {\scriptsize
EEM}}(\vec{p},\vec{p}\,',\lambda) \;\; ,
\label{meqfull} 
\end{equation}

\noindent 
where $e_q$ is the quark charge, 
$\lambda$ is the state of polarization of the photon, 
$\omega$ ($\vec{q}\,$) its energy (three-momentum) and $\vec{p}$
($\vec{p}\,'$) is the three-momentum of the initial (final) quark.  By
considering the usual electromagnetic current for point--like fermions
the single--quark transition
operator $O_{qq\gamma}^{\mbox {\scriptsize EEM}}$ reads:

\begin{equation}
O_{qq\gamma}^{\mbox {\scriptsize EEM}}(\vec{p},\vec{p}\,',\lambda)  
= \left(\frac{m}{E_p} \right)^{1/2}  \left(\frac{m}{E_{p'}} \right)^{1/2} \bar{u}(\vec{p}\,') \gamma_\mu u(\vec{p}\,)
\epsilon^{\mu \;
*}_\lambda (\vec{q}\,)  \;\; ,
\label{oqqeemfull}
\end{equation}
        
\noindent 
where $\epsilon_\lambda^{ \, \mu}(\vec{q}\,)$ is the photon
polarization four--vector, $E_p$ ($E_{p'}$) the on-shell energy of the
initial and final quarks ($E_p=\sqrt{m^2 + \vec{p}\,^2}$), 
and $m$ the mass of the quarks which is
assumed to be the same for all of them. 

        From (\ref{oqqeemfull}) the conventional way to derive a
non-relativistic transition operator is to proceed to a $(p/m)$
expansion keeping terms up to the first order
\cite{KONIUK80}.  Nevertheless for light baryons in a
quark model this procedure is under suspicion since the quarks move
inside the core with relativistic velocities and then $\langle p/m
\rangle$ can be even bigger than 1. On the other hand a more
reasonable expansion in terms of the relativistic velocity $(p/E)$ may
be slowly convergent since the value of $\langle p/E \rangle$ is
usually pretty close to 1.  Thus it seems more appropriate to consider
the whole relativistic operator Eq. (\ref{oqqeemfull}) in spite of the
fact that it is to be sandwiched between $3q$ baryon wave functions
obtained with a non-relativistic quark model. To this respect we
assume that once fitted the spectroscopy, the non-relativistic $3q$
wave function may emulate the relativistic one when relativistic
normalization and kinematical factors are considered
\cite{DESPLANQUES00}.

        The baryonic matrix elements for a process $B \longrightarrow
B' \gamma $ are easily computed 
from the single--quark matrix element Eq. (\ref{meqfull}) 

\begin{eqnarray}
\langle B' \gamma(\vec{q}, \lambda) | H | B \rangle 
 & = &\frac{3}{(2 \pi)^{3/2}} \frac{1}{(2 \omega)^{1/2}}
\delta^{(3)}(\vec{P} - \vec{P}\,' - \vec{q} \,) \nonumber \\
& & \int d\vec{p}_{\xi_{1}} \; \int d\vec{p}_{\xi_{2}} \; 
\Psi^*_{B'}(\vec{p}_{\xi_{1}},\vec{p}_{\xi_{2}} +
\sqrt{\frac{2}{3}} \vec{q}\:) \nonumber \\ 
 &  &  {\cal O}_{qq\gamma}(\vec{p}_3,\vec{p}_3\,',\lambda)
 \Psi_{B}(\vec{p}_{\xi_{1}},\vec{p}_{\xi_{2}}) \;\; ,
\label{mebgeneral}
\end{eqnarray}

\noindent where $\vec{P}$ ($\vec{P}\,'$) is the three-momentum of the
initial (final) baryon and  $\vec{p}_{\xi_1}$ and $\vec{p}_{\xi_2}$
are the
conjugate momenta of the Jacobi coordinates $\vec{\xi}_1$ and
$\vec{\xi}_2$. $\Psi$ stands for the
wave function of the baryons and the single--quark
transition operator has been particularized for the quark 3.

        All the dependence on a specific quark model for the baryons
is contained in the baryon wave functions. Hereforth we shall make use
of a spectroscopic potential model, very much detailed elsewhere
\cite{SILVESTRE85}, which contains,
aside from a linear confinement,  the 'minimal' one gluon exchange--like
terms. The explicit expression for the quark--quark  potential is

\begin{eqnarray}
\vI & = & \sum_{i<j} \frac{1}{2} \left[ \frac{r_{ij}}{a^2} - 
\frac{\kappa}{r_{ij}} + \frac{\kappa}{m_{i} m_{j}} 
\frac{\exp (-r_{ij}/r_{0})}{r_{0}^{2} r_{ij}} \vec{\sigma}_{i} 
\vec{\sigma}_{j}-D
\right] \;\; ,
\end{eqnarray}

\noindent with $a^2=1.063$ GeV$^{-1}$ fm, $\kappa=0.52$,
$r_0=0.4545$ fm and the quark mass is set to $m=0.337$ GeV. This
potential provides very good results for the spectroscopy of low--lying
baryons (ground states) as well as mesons. Concerning the excited states
the energies are reasonably predicted with the exception of the Roper
resonances. 

        To evaluate the electromagnetic form factor we have to consider the elastic $eN$ scattering process.
We take the Breit frame where $\omega=0$, $\vec{q}^{\; 2} =
Q^2$. We shall calculate the $N \rightarrow N\gamma$ amplitude from
(\ref{mebgeneral}) and extract the form factors from the corresponding
expression at the nucleonic level that reads:

\begin{eqnarray}
\label{macromebelec}
\langle N \gamma(\vec{q}, \lambda) | H_{NN\gamma} | N \rangle & =
&\frac{3 e}{(2 \pi)^{3/2}} \frac{1}{(2 \omega_{\gamma})^{1/2}}
\delta^{(3)}(\vec{P}_B - \vec{P}_{B'} - \vec{q}\,) \nonumber \\ & &
\hspace{-3cm} \chi_N' \left[\frac{G_E(Q^2)}{\sqrt{1+\frac{Q^2}{4 M^2}}}
\epsilon^{0\, *}_\lambda (\vec{q}\,) - i \frac{G_M(Q^2)}{2 M
\sqrt{1+\frac{Q^2}{4 M^2}}} (\vec{q} \times \vec{\sigma}_N) \cdot
\vec{\epsilon}\,^*_\lambda(\vec{q}\,) \right] \chi_N \;\; ,
\end{eqnarray}

\noindent  where
$\vec{\sigma}_N$ is the spin operator acting on the nucleon spinors
$\chi_N$, $\chi_{N'}$.

        Results for the electric and magnetic form factors are shown
in Fig. 2 (dashed lines) as compared to the conventional first-order
$(p/m)$ expansion (dash--dotted line) and to data. A look at the figures
shows significant discrepancies between the two calculations even for
low $Q^2$ values.

                Regarding the electric form factor the slope of
$G_E(Q^2)$ at the origin $Q^2 \longrightarrow 0$, is related to the
square mean charge radius of the nucleon. The $(p/m)$-expansion
neglects contributions to the radius coming from higher orders. These
contributions (Darwin-Foldy term) are present when the whole operator
is used giving rise to a bigger charge radius as compared to the $(p/m)$
value. Nonetheless in both cases it is still too small ( 0.238 fm$^2$ and
0.327 fm$^2$) as compared to data ($\langle r^2_p \rangle_{\mbox
{\tiny Exp.}} = 0.74 \pm 0.02$ fm$^2$ \cite{BROWN86}). 
This is a direct consequence of the
reduced size of the nucleon wave function which seems to be an
inevitable feature of any $3q$ model able to reasonably fit the
spectrum.
                
                        Concerning the magnetic form factor, the
magnetic moments calculated with (\ref{oqqeemfull}) are a 30 \% smaller than the ones obtained with
the $(p/m)$ expansion. The reason for this reduction is the
presence, for $Q^2 = 0$, of the energy factor $1/(2 E_3)$  
in the vector part of the quark current
instead of the mass factor $1/(2m)$. On the other hand the $Q^2$ 
dependence of this factor makes the magnetic
form factor go faster to zero when increasing $Q^2$ as compared to
the $(p/m)$ case.

        It is then clear the insufficiency of the EEM mechanism
        when combined with a spectroscopic quark model to explain
        the data, even if some relativistic kinematic corrections are
        included as in Eq. (\ref{oqqeemfull}).
        
\section{Dynamical relativistic corrections}

        Leaving aside for the moment kinematical corrections, we pay
        attention to dynamical relativistic corrections
        associated to the presence of quark-antiquark pairs in the
        baryonic medium. We certainly expect these
        corrections, that to some extent represent the effects of the mesonic
        cloud of the nucleon, to give sizeable contributions to the
        radius and to the magnetic moments as suggested by other
        approaches such as the relativistic chiral bag model.
        
                Following the ideas developed in a previous paper
\cite{CANO96} to treat strong pionic decays of baryons, we shall use
the $^3P_0$ quark pair creation as a way to generate the extra
$q\bar{q}$ pair in the baryonic medium. The $^3P_0$ operator written in a
relativistic form reads:

\begin{eqnarray}
H_{^3P_0} &  =  &
\beta \int d\vec{p} \left( \frac{m}{E_p}\right) \sum_{s,\tau,s',
\tau '} \left\{ \bar{u}_{s,\tau} (\vec{p}\,) v_{s',\tau '}(-\vec{p}\,) 
b_{s,\tau}(\vec{p}\,)  d^{\dag}_{s',\tau '}(-\vec{p}\,) \right. \nonumber \\
 &  & + \left.
\bar{v}_{s,\tau} (\vec{p}\,) u_{s',\tau '}(-\vec{p}\,) 
d_{s,\tau}(\vec{p}\,)  b^{\dag}_{s',\tau '}(-\vec{p}\,) \right\}
\label{3p0operator} \;\; ,
\end{eqnarray}

\noindent where $u$, $v$ stand for Dirac four--spinors and $b$, $d$ are
the usual annihilation quark and antiquark operators. $\beta$ is an
effective strength parameter that controls the pair formation in the
hadronic medium. From Eq. (\ref{3p0operator}) it is easy to check by
keeping terms up to $(\vec{p}/m)$ order that one can recover the
conventional non--relativistic  $^3P_0$ Hamiltonian \cite{LEYAOUANC73}.

 In this scenario, two
contributions can be considered.  First the recombined quark-antiquark
pair propagates in a resonant state which must be a vector meson in
order to have the photon quantum numbers (Fig. 1.b). At low momentum transfer
($Q^2 \lesssim$ 2-3 GeV$^2$) we can  restrict ourselves to the $\rho$ and
$\omega$ mesons. On the other hand, there is no reason to think that
this resonant contribution saturates the $| qqq \; q\bar{q} \rangle$
component and there could be
non-resonant propagation of the quark-antiquark pair as well (Fig. 1c).

{\bf 1. Resonant Diagrams}. 
The resonant amplitude is written from Fig. 1b. by considering
the two possible time orderings corresponding respectively to the
vector meson propagating forward and backward in time.

        For the electromagnetic vector meson-photon vertex we assume a
self--gauge invariant coupling $f_V F^{\mu \nu} V_{\mu \nu}$ that
guarantees that each time--ordered diagram is gauge invariant
separately.

        For the strong quark-antiquark vector meson vertex $\langle q
| H_{^3P_0} | q V \rangle$ we use the previously defined $^3P_0$
model, where the vector meson state is written in its relativistic
form:

\begin{eqnarray}
\label{vwffin}
|V (\vec{q}_V,\epsilon_V) \rangle & = & - \frac{1}{2}
\sum_{s,\tau,s',\tau'} \int \; d^3p_q \, d^3p_{\bar{q}} \Phi
\left(\frac{\vec{p}_q-\vec{p}_{\bar{q}}}{2}\right)
\delta(\vec{p}_q+\vec{p}_{\bar{q}}-\vec{q}_V) \nonumber \\ & &
\bar{u}_{s,\tau}(\vec{p}_q) \gamma_\mu  {\cal{O}}_\tau
v_{s',\tau'} (\vec{p}_{\bar{q}}) \epsilon_V^\mu(\vec{q}_V)
b^{\smdag}_{s,\tau}(\vec{p}_q)
d^{\smdag}_{s',\tau'}(\vec{p}_{\bar{q}}) |0\rangle \;\; ,
\label{vfinal}
\end{eqnarray}

\noindent where ${\cal{O}}_\tau$ fixes the isospin wave function of the
meson state ( ${\cal{O}}_\tau=\vec{\tau} (1)$ for an isovector
(isoscalar) meson). For the internal wave function $\Phi$ we have
taken a Gaussian form $\Phi(\vec{k}\,) = (R_V/\sqrt{\pi})^{3/2} \exp(-
k^2 R_V^2/2)$ whose parameter $R_V$ is fixed to the leptonic decay
width of the $\rho$ meson. Additional checks with a Coulombian wave
function shows that results are  little sensitive (less than 5 \%) 
to the choice of the
functional form of $\Phi$. Moreover, in the following we will assume
for the sake of simplicity the SU(3) relationship $f_\omega = 3
f_\rho$ and take for the $\rho$ and $\omega$ an averaged mass $m_V=
(m_\rho + m_\omega)/2$. The resulting single--quark transition operator
in the Breit frame ($\omega = 0$) is:

\begin{eqnarray}
O_{qq\gamma}^{\mbox {\scriptsize V}}(\vec{p},\vec{p}\,',\lambda) & = &
-e_q \frac{\beta}{f_V} \frac{E_V^{1/2}}{(2)^{3/2}}
\Phi^*_V(\frac{1}{2} (\vec{p}+\vec{p}\,')) \frac{1}{Q^2 + m_V^2}
\nonumber \\ & \times & 
\left\{ \epsilon^{0\;
*}_\lambda (\vec{q}\,) 
 \left( E_V - \frac{Q^2}{E_p+E_{p'}} \right) 
\left( \frac{\vec{\sigma} \cdot \vec{q} \; \vec{\sigma} \cdot
 \vec{p}}{E_{p}}- \frac{\vec{\sigma} \cdot \vec{p}\,' \; \vec{\sigma} \cdot
 \vec{q}}{E_{p'}} \right) \right. \nonumber \\ & - & \left. 
\vec{\epsilon}\,^*_\lambda (\vec{q}\,) \cdot \left[ 
\frac{\vec{\sigma} \cdot \vec{p}\,'}{E_{p'}} (Q^2 \, \vec{\sigma} -
(\vec{\sigma} \cdot \vec{q}\,)\, \vec{q}\,) +  (Q^2 \, \vec{\sigma} -
(\vec{\sigma} \cdot \vec{q}\,)\, \vec{q}\,) \frac{\vec{\sigma} \cdot
\vec{p}}{E_{p}}  \right]\rule{0pt}{17pt} \right\}.  
\nonumber \\  
\label{oqqvfull}
\end{eqnarray}

{\bf 2. Non--resonant diagrams}.
By proceeding in the same manner one can evaluate the matrix element
of the single--quark
transition operator for the non-resonant propagation of the
quark-antiquark pair. However, a difficulty immediately arises since
this operator, by its own, does not give rise to a conserved current.  
In order to see how gauge invariance
can be recovered it is necessary to understand the underlying physics in the $^3P_0$ operator. 

\section{Gauge invariant current}

        Let us assume that the $^3P_0$ operator is generated by some
residual interaction between quarks and gluons inside the baryon. This
residual interaction may be very complex and its detailed description 
would tantamount to unveil the structure of the hadronic
vacuum. Nonetheless, we shall show in a simplified model that the
generation of the pair from the vacuum (i.e. the use of the $^3P_0$
operator) must be accompanied by a mass renormalization. This mass
renormalization directly connected to the strength $\beta$ allows one
to restore gauge invariance. 

        The simplest modelization one can do of the residual
interaction is through the coupling of the quarks to a scalar mean
field $B(x)$:

\begin{equation}
{\cal L} = -g \bar{q}(x) B(x) q(x) \;\; ,
\label{lresidual}
\end{equation}

\noindent where $g$ is some unknown coupling constant. Four types of
diagrams come from Eq. (\ref{lresidual}). Two of them
correspond to the $^3P_0$ $q \bar{q}$ creation and 
annihilation under the identification $\beta = g B(x)$.
The other two diagrams give rise to a mass renormalization for
        quark and antiquark which can be written as $m= m_0 + g B(0)$.
In general the renormalized mass $m$
may be a very complicated function of $m_0$ and $g$, but since we are
interested in transition operators up to order $\beta$, it can be
reduced to the linear relationship quoted above. The important
fact is that the use of the $^3P_0$ model leads consistently to a mass
renormalization.

Now we are in conditions to understand how gauge invariance
can be recovered. The renormalization of the mass breaks the
conservation of the current associated to the EEM. Indeed, the
mass that appears in Eq. (\ref{oqqeemfull}) has to be interpreted as a bare
mass $m_0$ and as a consequence
the current associated to the EEM is not conserved anymore.
The breaking term is of the order $(E_p-E_{0 \, p})$, where
$E_{0 \, p}=\sqrt{m_0^2 + \vec{p}\,^2}$ is the energy corresponding to the
unrenormalized mass.  
However, the operators have to be written eventually in
terms of the
physical mass $m$ and therefore one has to replace $m_0$ by their
value in terms of the physical mass.  When doing so,
the terms that break gauge invariance in the EEM current and in 
the non-resonant sector cancel each other under the requirement:

\begin{equation}
m=m_0 + \beta/2 \;\; .
\end{equation}

        The resulting final single--quark transition operator, 
which respects gauge invariance, is
written as a sum of three terms:

\begin{equation}
O_{qq\gamma}(\vec{p},\vec{p}\,',\lambda)  = 
O_{qq\gamma}^{\mbox {\scriptsize EEM}}(\vec{p},\vec{p}\,',\lambda)  + 
O_{qq\gamma}^{\mbox {\scriptsize V}}(\vec{p},\vec{p}\,',\lambda)  +
O_{qq\gamma}^{\mbox {\scriptsize NR-EEM}}(\vec{p},\vec{p}\,',\lambda)
\;\; ,
\label{finaloperator}
\end{equation}

\noindent where $O_{qq\gamma}^{\mbox {\scriptsize EEM}}$ and
$O_{qq\gamma}^{\mbox {\scriptsize V}}$ are given by
Eqs. (\ref{oqqeemfull}) and (\ref{oqqvfull}) respectively being $m$
the physical mass and 

\begin{eqnarray}
O_{qq\gamma}^{\mbox {\scriptsize NR-EEM}}(\vec{p},\vec{p}\,',\lambda)
& = & e_q \frac{\beta}{8 \sqrt{E_p E_{p'} (E_p + m) (E_{p'} + m)}}
\vec{\epsilon}_\lambda\,^*(\vec{q}\,) \nonumber \\
& \times & \left[ i (\vec{\sigma} \times
\vec{p}\,') \left( \frac{\vec{p}\,^2}{E_p^2} + \frac{\vec{p} \cdot
\vec{p}\,'}{E_{p'}^2} \right) - i (\vec{\sigma} \times
\vec{p}\,) \left( \frac{\vec{p}\,'^{2}}{E_{p'}^2} + \frac{\vec{p} \cdot
\vec{p}\,'}{E_p^2} \right)\right. \nonumber \\
& - & i \vec{\sigma} \cdot (\vec{p}\,' \times \vec{p}) \left(
\frac{\vec{p}}{E_p^2} + \frac{\vec{p}\,'}{E_{p'}^2} \right)  + i  
(\vec{p} \times \vec{p}\,') \left( \frac{\vec{\sigma} \cdot
\vec{p}\,'}{E_{p'}^2} +\frac{\vec{\sigma} \cdot
\vec{p}}{E_p^2} \right) \left. \rule{0pt}{20pt} \right] . \nonumber \\
& & 
\label{oqqnonres}
\end{eqnarray}

\section{Results and discussion}

        The nucleon form factors obtained from our final gauge
invariant operator are shown in Fig. 2 (solid lines). The value of
the only free parameter $\beta$ has been chosen so that the magnetic
moment of the proton is fitted to its experimental value. The neutron
magnetic moment is also well reproduced, $\mu_n = -1.89$.

As a general
result we can say that in all cases the model represents an important
improvement with respect to the EEM predictions (dashed lines).
A comparison between
the two sets of curves gives a quantitative idea of the contribution
of the $| qqq \; q\bar{q}\rangle$ components to the electromagnetic
structure of the nucleon.

        The magnetic form factors are precisely reproduced up to
$Q^2=1.2$ GeV$^2$ (notice the $Q^4$ in the scale).
The main
(positive) contributions comes from the EEM, which gives up to 60 \%
at $Q^2=0$ (see Fig. 3). The
non-resonant contribution (NR-EEM) is also positive and it is very
relevant at low $Q^2$. In particular it gives almost a 40 \% of the
magnetic moment of the proton $\mu_p$. Finally, the resonant operator
contribution is negative and vanishes at $Q^2=0$, i.e. it gives 
no contribution
to the magnetic moments. As expected on general grounds,
the contribution of these mesonic components is important only around
the mass pole, though its absolute value in this model is still small
as compared to the non-resonant propagation. 

        Concerning the electric form factor, results differ from data
and follow the same trend in the whole range of $Q^2$ examined. The
predicted charge square mean radii ($\langle r_p^2\rangle = 0.48$
fm$^2$ and  $\langle r_n^2\rangle = -0.06$ fm$^2$) are still small as
compared to data ($\langle r_p^2\rangle = 0.74 \pm 0.02$
fm$^2$ \cite{BROWN86} and  $\langle r_n^2\rangle = -0.1215 \pm 0.0016$ fm$^2$
\cite{KOESTER76}). It is remarkable
that the resonant diagram accounts for 30 \% of $\langle r_p^2\rangle$
whereas the EEM diagram, that includes Darwin-Foldy terms and other
higher corrections gives $0.32$ fm$^2$. The 
lack of some contributions to the square
mean radius is manifest through the slope of
the curves at $Q^2$ which determines the difference with data at
higher $Q^2$. As can be seen from
Eq. (\ref{oqqnonres}), the non-resonant term does not take part in the
electric transitions since their time--like component vanishes.
Regarding the small value of the neutron charge radius, it
should be realized that in our case we have used SU(6) relations
($m_\rho = m_\omega$, $f_\omega = 3 f_\rho$) so that the diagonal
contributions of the SU(6) symmetric component of the wave function
(98.5 \% ) vanishes. The breaking of the former relations (coupling to
$2 \pi$ in one case, to $3 \pi$ in the other for instance) could
significantly change the charge radius of the neutron with very small
effects on the other form factors.
        
        The effective constant $\beta$ parametrizes, in the simplest
way, the creation of $q \bar{q}$ pairs in the hadronic medium. Its
effectiveness reflects gluon exchange interactions, maybe mostly
related to the confinement potential and other non-perturbative effects.
With respect to its particular value, and to 
ensure the consistency of the whole scheme, strong
processes such as pionic decays of resonances, and photo and
electroproduction of resonances  should be 
described within the same model and with the same value of
$\beta$ \cite{PROGRESS}. 
Therefore not much can be said about the reasonable value of
$\beta$ until this medium term program is carried out.

Our results can be compared with those of Ref. \cite{HELMINEN99},
where exchange currents and quark form factors contributions to the
charge radii are evaluated. The effects of exchange currents are
partially taken into
account in the $^3P_0$ model through the effective
value of $\beta$. However, $\rho$ and $\omega$ propagation diagrams in
our model incorporate quark--antiquark interactions and would thus 
be related to quark form factors rather than to
exchange currents.  

Finally, some comments are in order about the kinematical
relativistic corrections, which have been also the subject of recent
interest \cite{WAGENBRUNN00}. Undoubtedly these corrections are
important and affect the predicted values of the form factors by
introducing new $Q^2$ dependences which can also contribute
to the square charge radius. For example, the authors of
Ref.\cite{WAGENBRUNN00} obtain a value for $\langle r_n^2\rangle$
close to the experimental number. Nonetheless we do no think that
they could play, in an effective way,  the role of the dynamical
mechanisms described here: in particular concerning the contributions
of the $\rho$ and $\omega$ mesons to the nucleon form factors. As a
matter of fact one could use a simple counting of the normalization
and boost factors when employing relativistic wave functions instead
of non--relativistic ones \cite{DESPLANQUES00} without invoking the
detailed dynamics at all.

        In summary we have developed a consistent model to treat some
        relativistic dynamical corrections to the nucleon form
        factors. The model operator includes the effect of $| qqq \;
        q\bar{q} \rangle$ baryon components generated through a
        $^3P_0$ mechanism. Gauge invariance is ensured through the
        mass renormalization associated to such a mechanism. The only
        free parameter in the model is $\beta$, the strength of the
        creation of pairs in the hadronic medium, which is fitted to
        the magnetic moment of the proton. The results obtained
        seem to confirm our initial expectations about the
        consideration of $q\bar{q}$ contributions as an essential
        physical ingredient in the description of electromagnetic
        transition processes.

\vskip 1cm
This work has been supported in part by the EC-IHP
Network ESOP, contract HPRN-CT-2000-00130 and the DGESIC (Spain) 
under contract PB97-1401-C02-01.


\newpage

\begin{center}
{\bf Figure captions}
\end{center}

\begin{description}

\item{\bf Figure 1} (a) Elementary emission model: the photon is
emitted by one of the constituent quarks. (b) Resonant propagation of
a $q\bar{q}$ pair relevant for electromagnetic interactions.  (c)
Non-resonant propagation of a $q\bar{q}$ pair. In (b,c) the creation
of the extra $q\bar{q}$ pair is described by the $^3P_0$ model
(crosses).

\item{\bf Figure 2} Nucleon electromagnetic form factors 
calculated with the operator (\ref{finaloperator}) (solid line). Dashed
lines show the contribution of the EEM, Eq. (\ref{oqqeemfull}) and
dash--dotted lines correspond to the case where only the lowest
order in $(p/m)$ is retained in Eq. (\ref{oqqeemfull}).
For experimental data see \cite{GDATA}.

\item{\bf Figure 3} Relative contributions of the terms in
Eq. (\ref{finaloperator}) to the total value of $G_M^p/\mu_p$ (solid line).
The short--dashed line corresponds to the EEM term, the long--dashed
line to the V term and the  dash--dotted line to the  NR-EEM contribution.  

\end{description}

\newpage 
\
\vfill
\begin{center}
\psfig{file=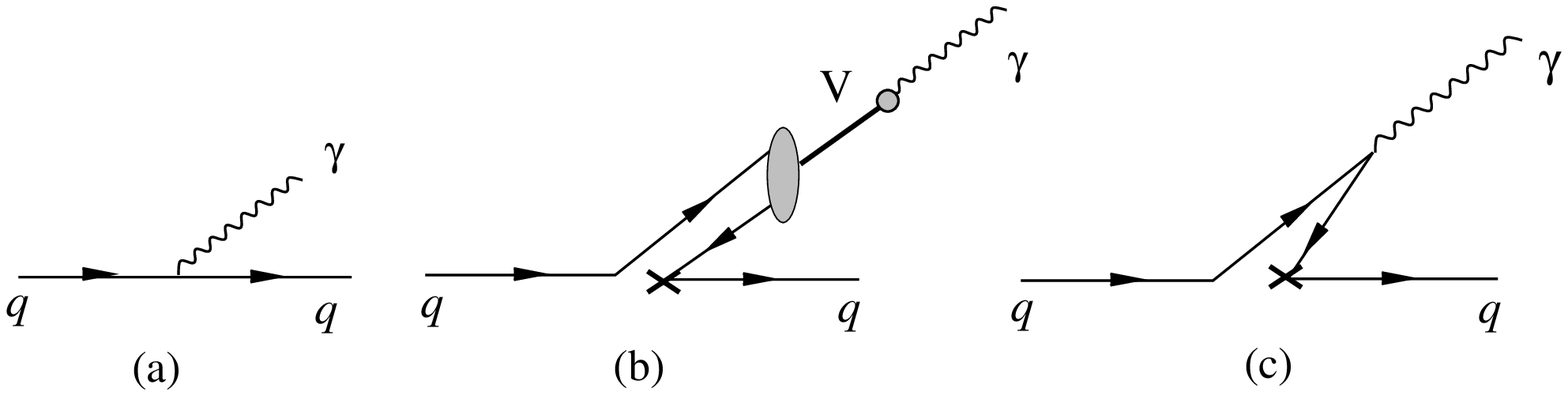,width=0.9\textwidth}
\vskip 1cm
{\bf Figure 1.}
\end{center}
\vfill
\newpage 
\
\vfill
\begin{center}
\psfig{file=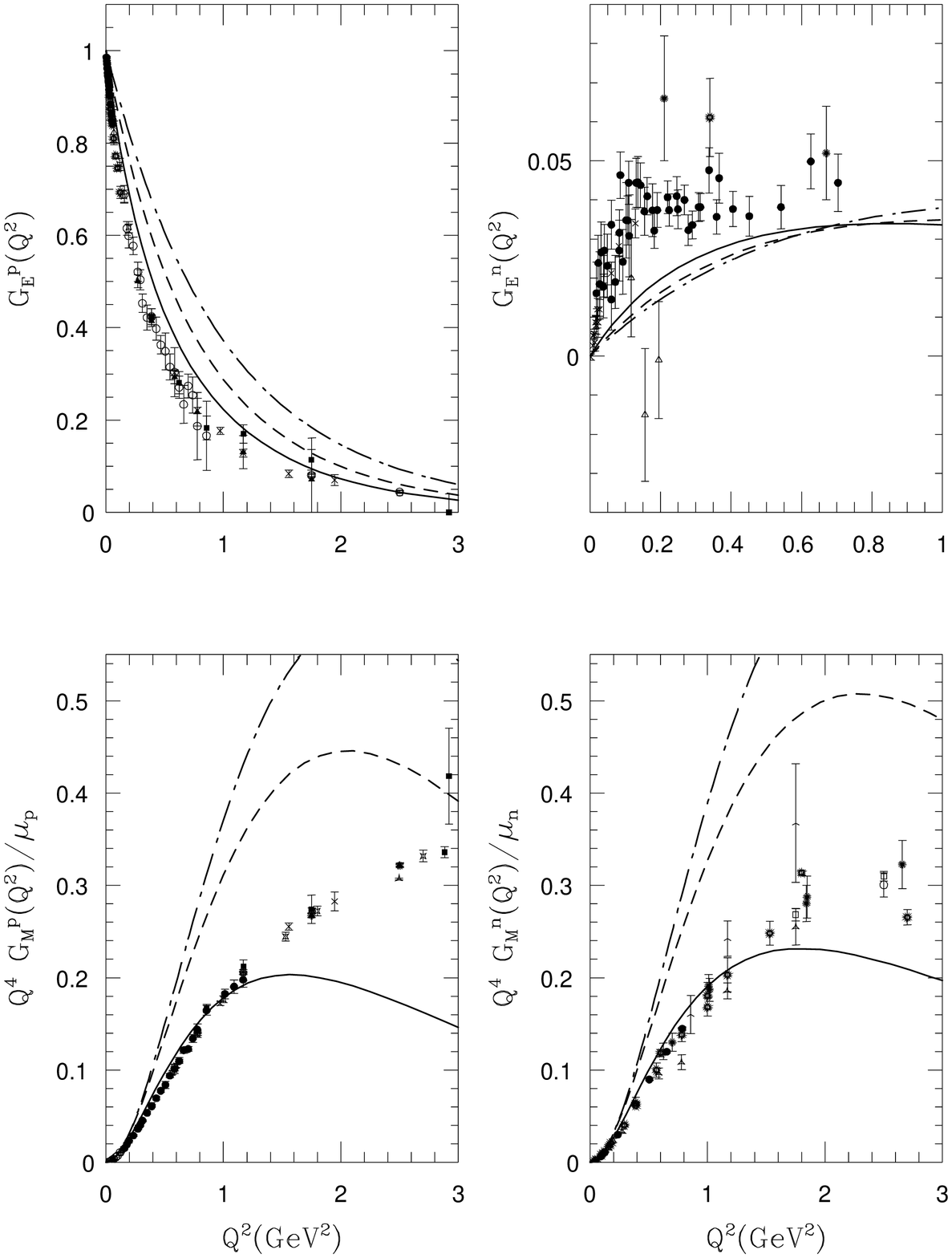,width=0.9\textwidth}
\vskip 1cm
{\bf Figure 2.}
\end{center}
\vfill
\newpage 
\
\vfill
\begin{center}
\centerline{\psfig{file=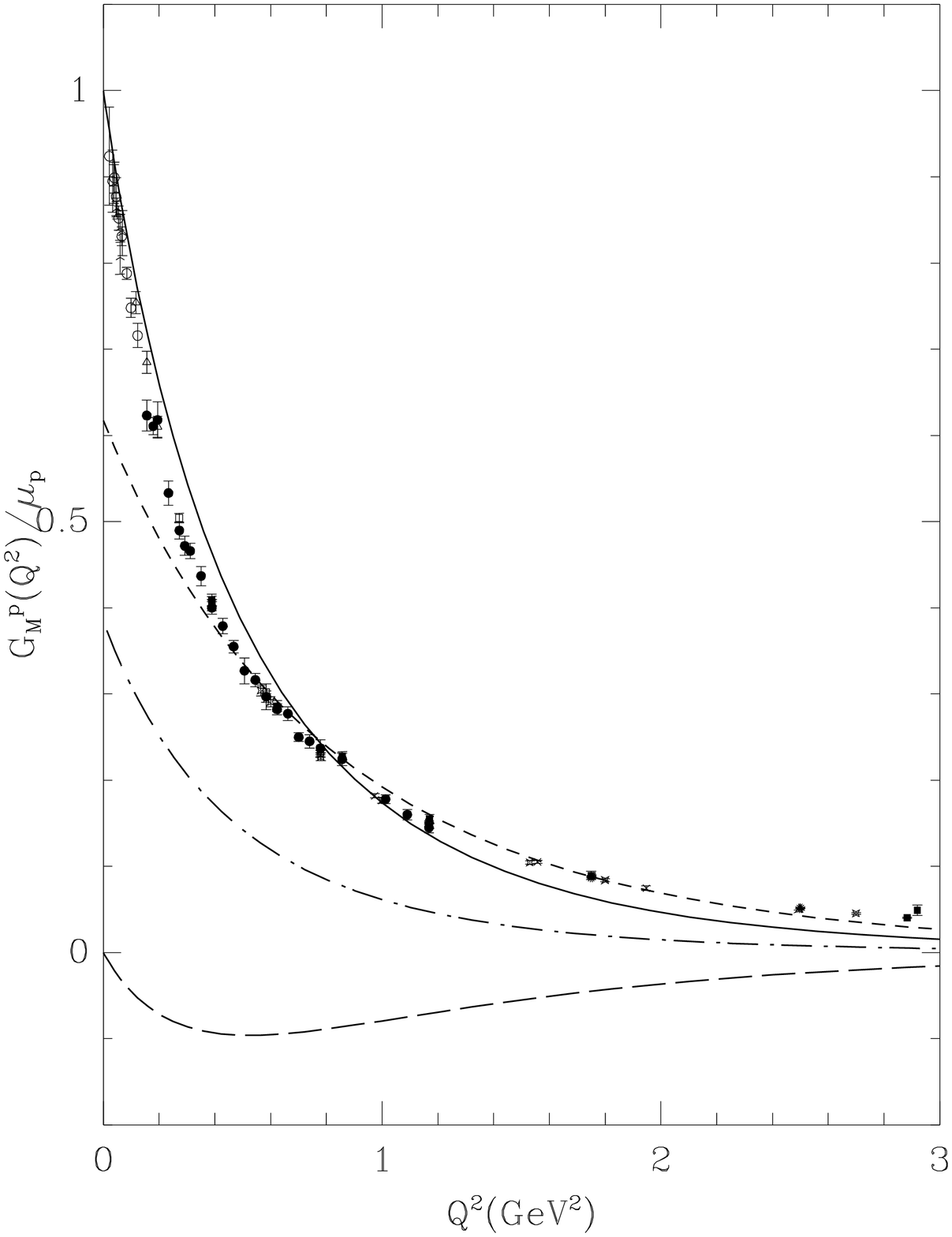,width=0.6\textwidth}}
\vskip 1cm
{\bf Figure 3.}
\end{center}
\vfill


\begin{thebibliography}{99}

\bibitem{BURKERT96} V.D. Burkert, Prog. Part. Nucl. Phys. 44 (2000)
273, and references therein; N. Isgur, nucl-th/0007008 (2000).

\bibitem{KONIUK80} R. Koniuk and N. Isgur, Phys. Rev. D21 (1980) 1868;  
F. Foster and G. Hughes, Rep. Prog. Phys. 46 (1983) 1445;
R. Sartor and P. Stancu, Phys. Rev. D33 (1986) 727;
M.M. Giannini, Rep. Prog. Phys. 54 (1991) 453;
R. Bijker, F. Iachello and A. Leviatan, Phys. Rev. C54 (1996) 1935;
M. Aiello, M.M. Giannini and E. Santopinto, J. Phys. G24 (1998) 753.

\bibitem{ROBSON93} D. Robson, Nucl. Phys. A560 (1993) 389; 
A.J. Buchmann, E. Hern\'andez and K. Yazaki, Nucl. Phys. A569 (1994)
661;  A.J. Buchmann, E. Hern\'andez and A. Faessler,
                        Phys. Rev. C55 (1997) 448;
E. Perazzi, M. Radici and S. Boffi, 
                        Nucl. Phys. A614 (1997) 346;
U. Meyer, A.J. Buchmann and A. Faessler,
                        Phys. Lett. B408 (1998) 19.

\bibitem{WARNS90} M. Warns, H. Schr\"oder, H. Rollnick, Phys. Rev. D42
                        (1990) 2215;
F.E. Close and Z. Li,   Phys. Rev. D42 (1990) 2194;
                        Z. Li and F.E. Close   Phys. Rev. D42 (1990) 2207; 
 S. Capstick,  Phys. Rev. D46 (1992) 1965, Phys. Rev. D46 (1992) 2864. 

\bibitem{DONG99} Y.B. Dong, K. Shimizu, A. Faessler and A.J. Buchmann,
Phys. Rev. C60 (1999) 035203.

\bibitem{KONEN90} W. Konen and H.J. Weber, Phys. Rev. D41 (1990) 2201;
S. Capstick and B.D. Keister, Phys. Rev. D51
                        (1995) 3598, nucl-th/9611055 (1996);
F. Cardarelli, E. Pace, G. Salm\`e and S. Simula,
                         Phys. Lett. B397 (1997) 13, Phys. Lett. B371
                        (1996) 7, Phys. Lett. B357 (1995) 267.

\bibitem{WAGENBRUNN00} R.F. Wagenbrunn, S. Boffi, W. Klink, W. Plessas
and M. Radici, Phys. Lett. B511 (2001) 33.

\bibitem{CANO00} F. Cano and M. Traini, Phys. Rev. C61 (2000) 065202.
 

\bibitem{CANO96} F. Cano, P. Gonz\'alez, S. Noguera and B. Desplanques,
                         Nucl. Phys. A603 (1996) 257;
F. Cano, P. Gonz\'alez, B. Desplanques and S. Noguera,
                         Z. Phys. A359 (1997) 315.

\bibitem{THEUSSL00}  L. Theussl, R.F. Wagenbrunn, 
B. Desplanques and W. Plessas, Nucl. Phys. A689 (2001) 394.

\bibitem{CANO98} F. Cano and  P. Gonz\'alez, Phys. Lett. B431 (1998) 270.

\bibitem{SILVESTRE85} B. Silvestre--Brac and C. Gignoux, Phys. Rev.
                        D32 (1985) 743.

\bibitem{DESPLANQUES00} B. Desplanques, B. Silvestre-Brac, F. Cano, P.
Gonz\'alez and S. Noguera, Few--Body Sys. 29 (2000) 169.

\bibitem{LEYAOUANC73} A. Le Yaouanc, L. Oliver, O. P\`ene and J.-C. Raynal,
                         Phys. Rev. D8  (1973) 2223, 
                         Phys. Rev. D11 (1975) 1272. 
                        
\bibitem{BROWN86} G.E. Brown, M. Rho and W. Weise, Nucl. Phys. A454
(1986) 669.


\bibitem{GDATA}   
        S. Platchkov {\it et al.}, Nucl. Phys. A510 (1990) 740;          
        D. Krupa {\it et al.}, J. Phys. G10 (1984) 455. 

\bibitem{KOESTER76} L. Koester, W. Nistler and W. Waschkowski,
Phys. Rev. Lett. 36 (1976) 1021.

\bibitem{PROGRESS} F. Cano, B. Desplanques, P. Gonz\'alez and
S. Noguera, in preparation. 

\bibitem{HELMINEN99} C. Helminen, Phys. Rev. C59 (1999) 2829.

\end{thebibliography}
\end{document}